\documentclass[iop,apj]{emulateapj}
\usepackage{verbatim}
\usepackage{latexsym}
\usepackage{amssymb}
\usepackage{amsmath}
\usepackage{longtable}
\usepackage{epsf}
\usepackage{graphicx}
\usepackage[colorlinks,urlcolor=blue,citecolor=blue,linkcolor=blue]{hyperref}

\def\Torb{T_{\rm orb}}
\def\re{R_{\rm E}}
\def\te{t_{\rm E}}

\shorttitle{Periodic Signals in Binary Microlensing Events}
\shortauthors{Guo et al.}

\begin{document}

\title{Periodic Signals in Binary Microlensing Events}

\author{Xinyi Guo$^{1,2}$, Ann Esin$^3$, Rosanne Di\thinspace Stefano$^1$ and Jeffrey Taylor$^3$}
\affil{$^1$Harvard-Smithsonian Center for Astrophysics, 60 Garden Street, Cambridge, MA 02138, USA \\
$^2$Department of Physics and Astronomy, Pomona College, 610 North College Avenue, Claremont, CA 91711, USA \\
$^3$Department of Physics, Harvey Mudd College, 301 Platt Blvd., Claremont, CA 91711, USA}

\begin{abstract}
Gravitational microlensing events are powerful tools for the study of stellar populations. In particular, they can be
used to discover and study a variety of binary systems. A large number of binary lenses have already been found
through microlensing surveys and a few of these systems show strong evidence of orbital motion on the timescale
of the lensing event. We expect that more binary lenses of this kind will be detected in the future. For binaries
whose orbital period is comparable to the event duration, the orbital motion can cause the lensing signal to deviate
drastically from that of a static binary lens. The most striking property of such light curves is the presence of quasi-
periodic features, which are produced as the source traverses the same regions in the rotating lens plane. These
repeating features contain information about the orbital period of the lens. If this period can be extracted, then
much can be learned about the lensing system even without performing time-consuming, detailed light curve
modeling. However, the relative transverse motion between the source and the lens significantly complicates the
problem of period extraction. To resolve this difficulty, we present a modification of the standard Lomb–Scargle
periodogram analysis. We test our method for four representative binary lens systems and demonstrate its
efficiency in correctly extracting binary orbital periods.
\end{abstract}

\maketitle

\section{Introduction}
Over the past two decades gravitational microlensing surveys have
transformed from a bold theoretical idea \citep{Pacz1986} into an
important tool for studying the stellar population of our
Galaxy. Following Paczy\'{n}ski's proposal, four different monitoring
programs, MACHO \citep{Alcock1997}, OGLE \citep{Udalski1993}, EROS \citep{Aubourg1993} and 
MOA \citep{Yock1998}, for years have collected data on the Galactic Center as
well as the Large Megellanic Cloud and the Small Magellanic Cloud, detecting many thousands of microlensing events.
Two microlensing surveys that are still currently in operation,
OGLE-IV and MOA, continue to add to this number at a rate of roughly
2000 events per year.  While the majority of the detected events are
consistent with being produced by a single-mass lens, a sizable
fraction shows clear evidence of lens binarity \citep{Alcock2000, Jaro2004, Jaro2010}. 
These are
particularly interesting, because more complicated lightcurves 
\citep[see, e.g.][]{Mao1991} produced by binary lenses allow us to learn
considerably more about these systems \citep[see, e.g.][]{Mao1995, DiStefano1997}.

The signatures of lens binarity are most prominent when the projected orbital 
separation between the two binary components, $a$, is of the order of the 
Einstein radius: 
\begin{equation}
\re=\sqrt{\frac{4 G M}{c^2}\frac{(D_{\rm s}-D_{\rm l})D_{\rm l}}{D_{\rm s}}},
\end{equation}
where $D_{\rm s}$ and $D_{\rm l}$ are the distances to the source and the lens 
respectively, and $M$ is the total 
mass of the lens.  The timescale for the duration of the lensing event is 
determined by the Einstein radius crossing time 
\begin{equation}
\te=\re/v,
\end{equation}
where $v$ is the relative transverse velocity between the lens and the source as
viewed by the observer.  For solar-mass lenses in the Galactic disk
$\te$ is typically on the order of months, while the orbital period of
the lensing binary, $\Torb$, is more likely to be measured in
years.  As a result, most binary lensing events can be modeled while
completely neglecting the orbital motion of the two masses 
\citep[see, e.g. recent estimates by][]{Penny2011a}.  
Nevertheless, a handful of
microlensing light curves do show convincing evidence of binary phase
change during the event \citep{Dominik1998, Albrow2000, An2002, Jaro2005, Hwang2010, Ryu2010, 2013ApJ...778..134P,Shv2014}.  
Recently, there has been increased interest in the possibility of finding lensing events for
which the timescale ratio, defined as 
\begin{equation}
{\cal R} = \te/\Torb, 
\end{equation}
is of order the of or greater than unity \citep{Penny2011b,DSE14,Nucita2014}.
{In addition, new techniques for finding such systems using astrometric
microlensing are being developed \citep{2014MNRAS.439.3007S}.}

If the orbital period can be determined for such systems, then it will
provide a new avenue for extracting the lens parameters. Specifically,
the period relates the total mass to the orbital separation. If the
mass ratio of the binary and their separation in units of the Einstein radius can also be determined, then we will have some of the
key elements needed to specify the binary. In cases when some
additional information is available, e.g., because the lens is located nearby
and/or because parallax or finite-source-size effects are detected, a
full binary solution may be obtained.  Unlike single-lens light curves
which have a simple analytical form \citep{Pacz1986}, binary lenses can
produce very complex magnification patterns that cannot be expressed
in a closed form.  Even for a stationary binary lens, the process of
fitting an observed light curve can be complex \citep{DiStefano1996}.
The addition of orbital motion complicates the fitting process even
further.  Fortunately, in the limit where ${\cal R} \gtrsim 1$ the binarity
features observed in the microlensing light curves show very strong
periodicity which allows $T_{\rm orb}$ to be determined via simple
Lomb-Scargle (LS) periodogram analysis \citep{DSE14}.  In this paper,
however, we are primarily interested in the regime where ${\cal R} \lesssim 1$.  
For these longer-period binaries, the interplay between
the orbital motion and the projected motion of the source causes the
period determined via LS analysis to differ significantly from $T_{\rm
  orb}$ \citep{DSE14}.  
Here, we describe a modified timing analysis
method which allows us to compensate for this relative motion and
extract the correct binary period of the lens. 
     
\section{Microlensing by a Rotating Binary Lens} 
\label{sec:parameters}

A microlensing event produced by a single lens can be completely
characterized by three parameters: $\re$, the Einstein radius of the
system; $b$, the distance of closest approach between the lens and the
projection of the source onto the lens plane (measured in units of
$\re$); and $v$, the transverse velocity of the source with respect to
the lens.  In general, we can express the separation between the lens
and the source projection (measured again in units of $\re$) as
\begin{equation}\label{eq:u}
u(t)=\sqrt{b^2+[(t-t_0)/\te]^2},
\end{equation}
where $t_0$ is the time of closest approach.  The resulting light curve
is described in terms of the amplification function, defined as the
ratio of the observed lensed and unlensed fluxes, which takes the form
\citep[e.g.][]{Pacz1986}
\begin{equation}\label{eq:amp}
A_{\rm SL}(t)=\frac{u^2+2}{u\sqrt{u^2+4}}.
\end{equation}

\begin{figure}[h]
\begin{center}
\includegraphics[width=3in]{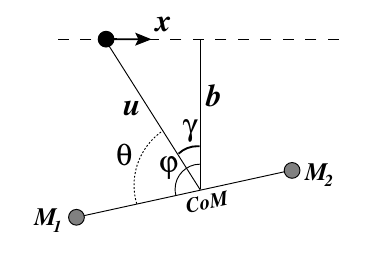}
\end{center}
\caption{Geometry of the lensing system.  The black dot shows the
  projection of the source location onto the lens plane.  The
  trajectory of the source in the lens plane is given by the dashed
  line.  Note that we refer to the vertical line connecting the center
  of mass of the binary lens and the point of closest approach of the
  source as a CoM-b-axis.}
\label{fig:geometry}
\end{figure}

When the lens is a binary, in addition to the total mass of the
system, $M=M_1+M_2$, we need to specify the mass ratio $q=M_1/M_2$ and
the dimensionless semi-major axis of the binary, $\alpha=a/\re$, to
fully describe the microlensing event.  For simplicity, in this paper,
we consider only face-on, circular binary orbits so that $\alpha$ is equal
to the binary separation at all times.  Since the lens is rotating,
two more parameters are necessary: the value of the binary phase angle
(shown as angle $\varphi$ in Fig. \ref{fig:geometry}) at the moment of
closest approach, $\varphi_0$, and the direction of binary rotation with
respect to the relative motion vector.  For the source velocity shown
in Fig. \ref{fig:geometry}, we define a binary to be {\it prograde} if
it is rotating clockwise and {\it retrograde} otherwise.

Since general relativity is a nonlinear theory, the light curve
produced by a binary lens is more complicated than a simple
superposition of two amplification functions and cannot be described
by an analytical formula.  However, we can calculate the amplification value
numerically as a function of source position in the lens plane
\citep{Schneider1986, Witt1990}.  The result depends only
on $\alpha$ and $q$.  Given $b$ and $v$, we can calculate the trajectory of
the source projection in the plane of the binary lens and construct 
the resulting light curve.   

\begin{figure}[h]
\begin{center}
\includegraphics[height=3in]{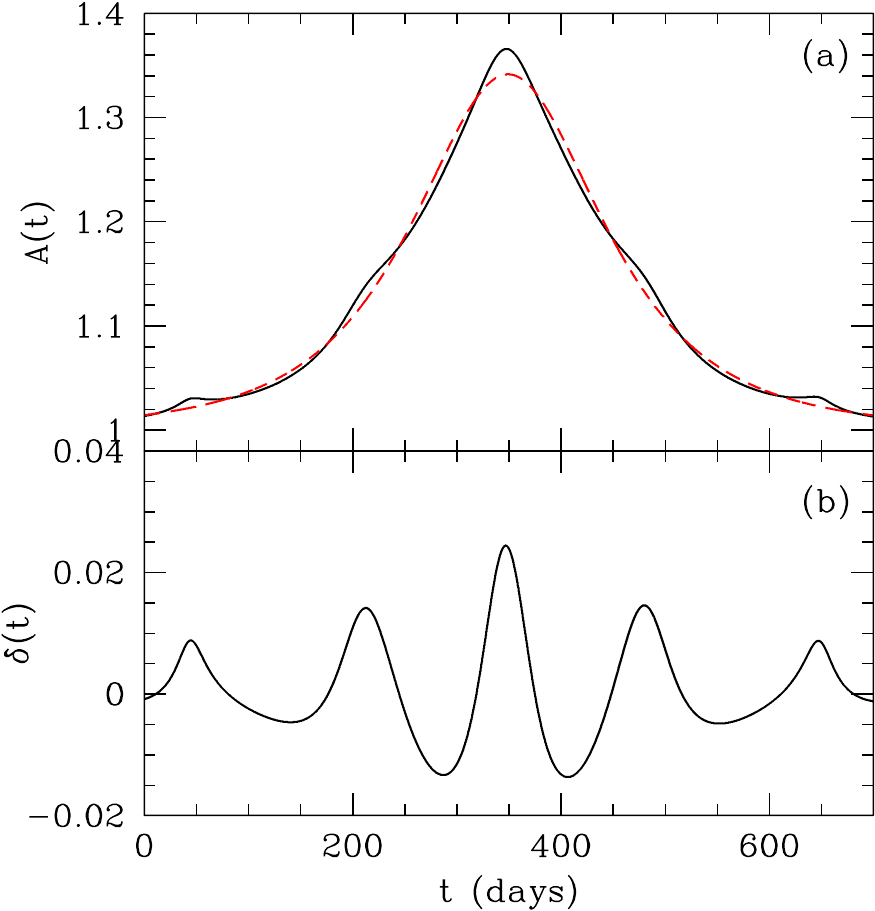}
\end{center}
\caption{(a) The black solid line shows a sample theoretical light
  curve with $b=1.0$ produced by our standard binary lensing system
  (see Section \ref{sec:stdsys}) rotating in retrograde direction.  The
  dashed red line shows the least-$\chi^2$ fit single-lens light curve
  $A_{\rm SL}(t)$.  (b) The residual light curve, defined as
  $\delta(t)=A(t)-A_{\rm SL}(t)$, clearly shows the effects of lens
  binarity. The oscillatory pattern contains information about the
  orbital motion of the lens.}
\label{fig:samplelc}
\end{figure}

Fig. \ref{fig:samplelc}(a) shows a sample light curve produced by our
standard binary lens system which consists of two $0.5 M_\odot$ stars (i.e. $M=1M_\odot$, $q=1$)
in a circular orbit with $a=1\rm AU$. The lens-source system is set up (described in detail in
Section \ref{sec:stdsys}) so that $\alpha=0.25$ and the timescale ratio is $\mathcal{R}=0.31$, 
representative of the $R\lesssim 1$ regime we are interested in.  
Over-plotted as a dashed red line is the best-fit
single-lens light curve (Eq. \ref{eq:amp}).  The overall shapes of the two
light curves are very similar, which is not surprising for a binary with
a fairly low value of $\alpha = 0.25$.  However, subtracting the
single-lens fit from the binary light curve reveals a clear oscillatory
pattern in the residuals, as shown in Fig. \ref{fig:samplelc}(b).

\begin{figure*}[tbp]
\begin{center}
\includegraphics[width=5.3in]{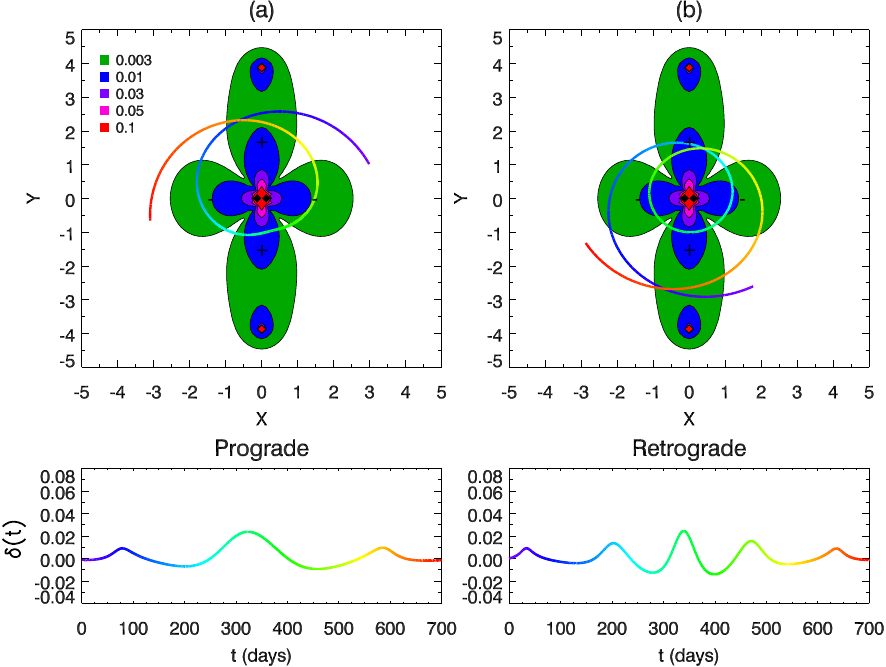}
\caption{\label{fig:proretromaglc} Both panels show the contour plot
  of the absolute residual magnification function,
  $|\delta(x,y)|=|A(x,y)-A_{\rm SL}(x,y)|$, produced by the standard
  system (described in Section \ref{sec:stdsys}). The contour map is
  centered on the center of mass of the binary lens and the axes are
  in units of $\re$. The two binary components are located on the
  $x$-axis at $x=\pm \alpha/2 = \pm 0.125$, indicated by the black dots. The horizontal elongated petals
  correspond to negative residuals and vertical petals correspond to
  positive residuals. The legend in panel (a) specifies the minimum
  values of $|\delta(x,y)|$ that correspond to different contour
  colors.  As the source moves past the lens,
  its trajectory in the lens plane spirals in and out, producing the
  oscillatory residual pattern seen in the lower panels.  Note that
  the colors of the trajectories and the residual curves are matched.
  Panel (a) shows the trajectory of the source with $b=1.0$ for the
  prograde binary lens, while panel (b) corresponds to the retrograde
  case (the light curve is identical to the one shown in
  Fig. \ref{fig:samplelc}).}
\end{center}
\end{figure*}

To understand the relationship between this periodic pattern in the
residuals and the orbital motion of the lens, it is instructive to
examine the trajectory of the source in the lensing plane.  In
Fig. \ref{fig:proretromaglc} we show this trajectory superimposed onto
the contour plot of the absolute value of the residual magnification
$|\delta(x,y)|=|A(x,y)-A_{\rm SL}(x,y)|$.  Here $x$ and $y$ are
coordinates in the lens frame, scaled to $\re$.  The residual contours
exhibit a nearly-symmetrical four-lobe shape. Note that the longer
lobes (lined up with the y-axis) correspond to positive $\delta(x,y)$
values , while the shorter ones have $\delta(x,y) < 0$.  For face-on
circular orbits $\alpha$ is time-independent and the amplification map 
is static.  In the co-moving and co-rotating frame of the
lens, the source appears to spiral in as it moves towards the point of
closest approach and spiral out as it moves away from the lens.  If
the source were stationary, one full rotation of the binary lens would
result in two full oscillations of the residual light curve. Thus, the
periodicity of the residual signal is related to the orbital period
of the binary lens by $T \sim \Torb/2.$ Note that the
relation is not exact due to the modification caused by relative
transverse motion between the source and the lens.

To get a quantitative understanding of the interplay between source
motion and binary rotation, we need to derive an equation for the time
evolution of angle $\theta$, as shown in Fig. \ref{fig:geometry}.
Assuming that the source moves with velocity $v$ in the lens plane,
its distance to the CoM-b-axis, scaled to $\re$, takes the form
\begin{equation}
x=\frac{t-t_0}{\te}.
\end{equation}
The angle from the CoM-b-axis to the source is then given by
\begin{equation}\label{eq:gamma}
\gamma(t)=-\arctan \frac{x}{b} = -\arctan \frac{t-t_0}{b \te}
\end{equation}
as shown in the geometry of the lensing system (Fig. \ref{fig:geometry}).
Notice that $\gamma(t)>0$ when $t<t_0$ and $\gamma(t)<0$ when $t>t_0.$
The phase angle from CoM-b-axis to the axis of the binary system is
\begin{equation}\label{eq:phi(t)}
\varphi(t)=\varphi_0\mp \omega_{orb} t,
\end{equation}
where $\omega_{\rm orb}=\frac{2\pi}{T_{\rm orb}}$ is defined as the orbital
frequency of the binary lens and ``$-$'' sign is adopted for prograde motion while ``$+$'' sign
corresponds to retrograde motion.  Finally, the angle between the source position and
the axis of the binary is then
\begin{align}
\theta(t)&=\varphi(t)-\gamma(t) \nonumber \\
&=
\phi_0\mp \omega_{\rm orb}t + \arctan \frac{t-t_0}{b \te}.
\label{eq:theta}
\end{align}
The distance from the source to the center of mass in the lens plane, $u(t)$
is given by Eq. \eqref{eq:u}.
Note that the position of the source in the co-rotating lens plane is simply 
\begin{equation}
\label{eq:XY}
x(t)=-u(t)\cos\theta(t),\qquad y(t)=u(t)\sin\theta(t).
\end{equation}

A natural consequence of Eq. \eqref{eq:theta} is that the angular
velocity of the source in the lens plane, $d{\theta}/dt$, is smaller
for a prograde system than that for a retrograde system. This effect
manifests itself clearly in the apparent period of residual signals
shown in Fig. \ref{fig:proretromaglc}, which compares the trajectories
of the source for a prograde and retrograde system with otherwise
identical physical parameters. The retrograde residual light curve
shows much faster oscillations because it has a larger value of
$d{\theta}/dt$.

\section{Timing Analysis}

We now address the question of how to extract the 
information about the orbital periodicity of the binary lens
from the observed microlensing light curves.  
We start by generating a theoretical binary light curve, add noise to 
simulate real survey data, subtract the dominant single-lens
signal, perform spectral analysis on the
the residuals, and examine the accuracy of the inferred orbital
period.

\subsection{Modeling the Observations}
We can generate a theoretical light curve of arbitrary duration.  To
approximate realistic observing conditions, we take the photometric
uncertainty to be $1\%$ for most of our calculations. (We discuss the
results for high-precision photometry in Section \ref{sec:lownoise}.)  At $u
= 3$, Eq. \eqref{eq:amp} gives $A_{\rm SL}\sim 1.01$, and so the overall
amplification level is marginally distinguishable from random noise
fluctuation.  Based on this, we define the span of each
lensing event to be from $3 \te$ before the time of closest approach
to $3 \te$ after the closest approach.  For all of our light curves we
assume a fairly conservative average sampling interval of $\Delta t=
2\; {\rm days}$.\footnote{In fact, the OGLE-IV program monitors a significant portion of the 
Galactic Bulge with $\Delta t< 2$ days (\url{http://ogle.astrouw.edu.pl/sky/ogle4-BLG/}) and the upcoming Korean Microlensing Telescope Network will achieve even higher cadence with its wider longitude coverage \citep{Henderson14}. }  With this average rate, we sample the theoretical
light curve at random times $\{t_j\}$ and add random Gaussian noise
with mean 0 and standard deviation $0.01$ to each data point.

Finally, we fit Eq. (\ref{eq:amp}) to our simulated light curve, 
both to obtain the residual light curve $\delta(t)$ as well as to determine 
our best-guess values for $b,t_0,
\te$, which are necessary for the subsequent timing analysis.
We found that the fitting procedure works much better after smoothing out
the sharp features, if any, produced by caustic crossings.

\subsection{Modified Lomb-Scargle Analysis}
Once we have  the residual light curve we are ready to perform spectral
analysis to extract the underlying periodicity of the binary.  
We base our period extraction method on the classical 
LS periodogram analysis \citep{Scargle1982}, a modification of the
Fourier transform designed for unevenly sampled data that has the
advantage of time-translation invariance.

\begin{figure}
\begin{center}
\includegraphics[width=3in]{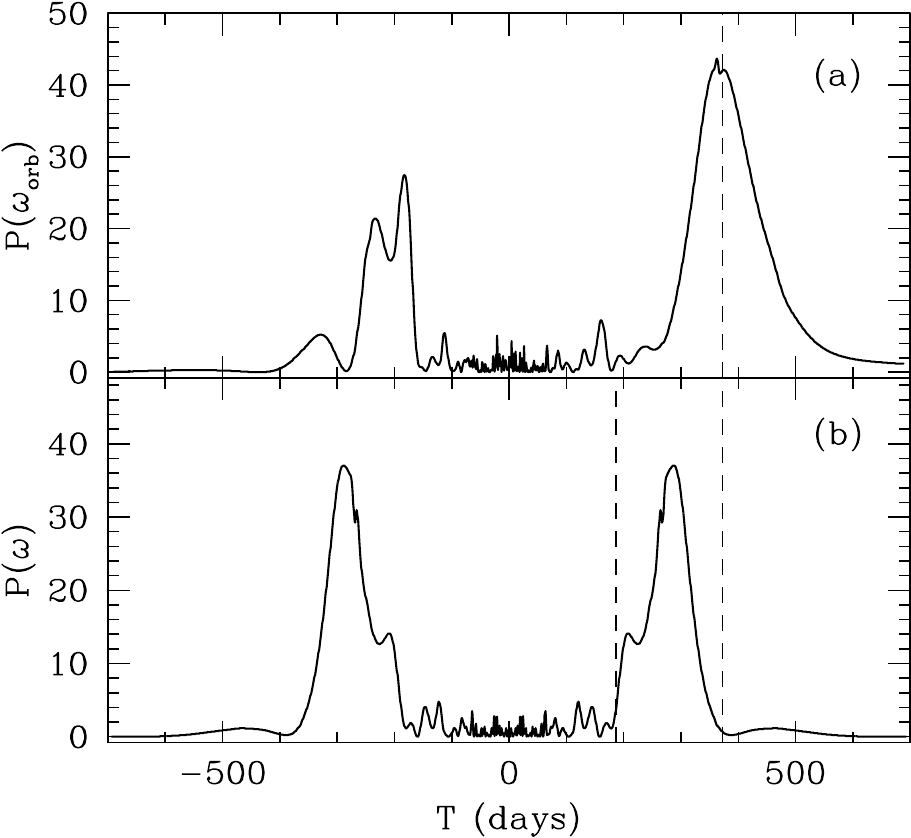}
\caption{Periodogram of the light curve shown in Fig. \ref{fig:samplelc}.  
 Panel (a) shows the results
  produced by the modified LS analysis.  The highest-power peak
  corresponds to a period of 362\,days, which is very close to the actual
  orbital period of the binary shown with a dashed line.  Panel (b)
  displays the results of the standard LS periodogram analysis (here
  the power spectrum for negative period values is simply a mirror
  image of the positive part).  The best-fit period of 266\,days is
  clearly very different from either $\Torb$ or $\Torb/2$ shown by
  dashed lines.}
\label{fig:oldnewpowerspectrum}
\end{center}
\end{figure}

For time series $\{\delta(t_j)\}$ with $N$ data points, the
LS periodogram as a function of frequency $\omega$ is
defined as
\begin{equation}
\label{eq:LS}
P(\omega)=\frac{1}{2}
\left[
\frac{\left(\sum_{j=1}^N \delta(t_j) C_j\right)^2}{\sum_{j=1}^N C_j^2}+
\frac{\left(\sum_{j=1}^N \delta(t_j) S_j\right)^2}{\sum_{j=1}^N S_j^2}
\right]
\end{equation}
where $C_j$ and $S_j$ are given by
\begin{equation}\label{eq:Cj}
C_j=\cos(\omega(t_j-\tau)),
\end{equation}
\begin{equation}\label{eq:Sj}
S_j=\sin(\omega(t_j-\tau)),
\end{equation}
and $\tau$ is defined as
\begin{equation}\label{eq:tau}
\tau =\frac{1}{2\omega} \tan^{-1}
\left[
\frac{\displaystyle\sum_{j=1}^N\sin 2\omega t_j}
{\displaystyle\sum_{j=1}^N\cos 2\omega t_j}
\right]
\end{equation}

Unsurprisingly, the unmodified LS periodogram analysis does not 
generally yield the correct binary period, since the frequency of the
observed oscillatory signal, $\omega$, is affected by the relative
source-lens motion (see Fig. \ref{fig:proretromaglc}) and is therefore
not equal to the the binary orbital frequency, $\omega_{\rm orb}$.
Figure \ref{fig:oldnewpowerspectrum}(b) shows the LS power spectrum
for a microlensing light curve with $b=0.4$ generated by our standard
lensing system.  The peak power corresponds to a period of 266\,days, 
which is very different from both the true orbital period of the system,
$T_{\rm orb} = 372.5$\,days, as well as $\Torb/2$ (which we expect
to come closer to reflecting the true variability timescale of the
lensing signal due to the bi-fold symmetry of the magnification map produced
by this binary, as illustrated by Fig. \ref{fig:proretromaglc}).

In Section \ref{sec:parameters} we argued that the progress of the lensing event 
is governed by the time evolution of angle $\theta$, defined by Eq. 
(\ref{eq:theta}), which takes into account both binary rotation and 
source motion.  Thus, we propose to modify the oscillating coefficients
$C_j$ and $S_j$, essentially replacing $\omega t_j$ with $\theta(t_j)$:  
\begin{equation}\label{eq:modCj}
C_j=\cos[2\{\omega_{\rm orb}(t_j-\tau)-\gamma(t_j)\}], 
\end{equation}
\begin{equation}\label{eq:modSj}
S_j=\sin[2\{\omega_{\rm orb}(t_j-\tau)-\gamma(t_j)\}].
\end{equation}
We dropped the $\varphi_0$ term in Eq. \eqref{eq:theta} since it simply
produces a time translation of the signal without affecting the
results of the LS analysis.  The extra factor of 2 originates from the
fact that the binary magnification pattern repeats itself twice over
one binary revolution, and so the period directly observed in the data
would be roughly half of the actual binary period, as we already
pointed out above.  Note that the definition of $\tau$ remains the
same apart from replacing $\omega$ with $\omega_{\rm orb}$ in
Eq. (\ref{eq:tau}).  Finally, the approximate values of $b$, $t_0$ and
$\te$ necessary for evaluating $\gamma$ were obtained when fitting
Eq. (\ref{eq:amp}) to the original light curve.

Unlike the standard LS periodogram analysis for which only positive 
$\omega$ values have meaning,  $\omega_{\rm orb}$ can have either sign.
Following the sign convention in Eq. \eqref{eq:theta}, when interpreting 
the results of the modified periodogram,
$\omega_{\rm orb} < 0$ corresponds to prograde systems while 
$\omega_{\rm orb} > 0$ corresponds to retrograde systems.

Fig. \ref{fig:oldnewpowerspectrum} illustrates the difference between
the results of the standard (panel (b)) and modified (panel (a))
periodogram analyses.  Our new method produces a sharp power 
peak corresponding to the period $T = 2 \pi/\omega_{\rm orb} = 362$\,days, 
within 3\% of the actual binary period.  In addition, a positive period 
value correctly identifies the lens as a retrograde binary.

\subsection{False Alarm Probability}

The final issue we need to address is how to 
determine the minimum power level, $P_{\rm cut}$, such that any peak
with a power above $P_{\rm cut}$ is unlikely to be produced by
chance. \cite{Scargle1982} derived a formula relating the false alarm
probability $p_{\rm f}$ and $P_{\rm cut}$:
\begin{equation}
\label{eq:Pcut}
P_{\rm cut}=-\ln [1-(1-p_f)^{1/N_{\omega}}],
\end{equation}
where $N_{\omega}$ is the number of independent frequencies searched.
For $p_f \ll 1$, we can approximate Eq. \eqref{eq:Pcut} as
\begin{equation}
\label{eq:Pmin}
P_{\rm cut}\approx \ln(N_{\omega}/p_f)=\ln(N_{\omega})+\ln(1/p_f).
\end{equation}
For a given value of $p_f$, $P_{\rm cut}$ depends only on
$N_{\omega}$.  There is no clear consensus in the literature on how to
calculate $N_{\omega}$ for a given set of data, since the precise
answer depends on the sampling method \citep{Gilliland1985, Baliunas1985,
Horne1986}. However, it is clear that $N_{\omega}$ must be on the order
of the number of data points, $N$. We can thus rewrite
Eq. \eqref{eq:Pmin} as
\begin{equation}
P_{\rm cut}-\ln(N)=\ln(N_{\omega}/N)+\ln(1/p_f),
\end{equation}
and determine the term $\ln{N_{\omega}/N}$ empirically, using the same
sampling scheme as we do for our light curves.

\begin{figure}[h]
\begin{center}
\includegraphics[width=2.5in]{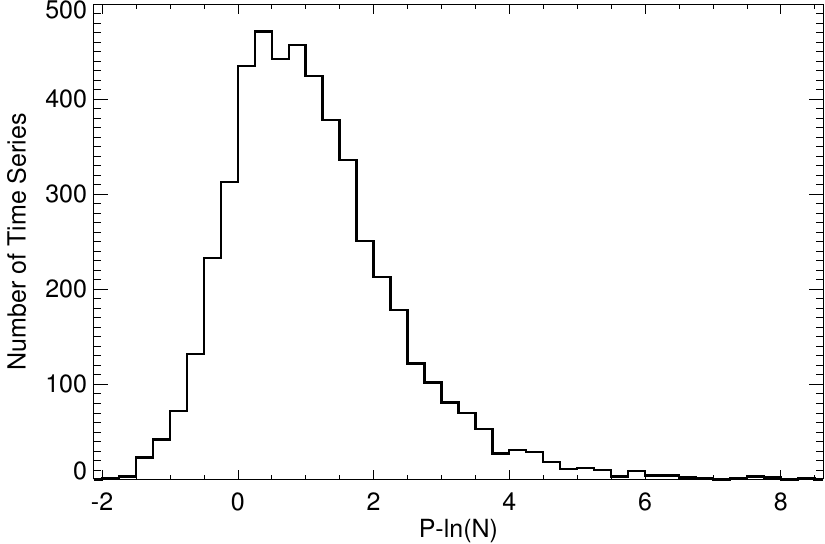}
\end{center}
\caption{Distribution of $P_{\rm max}-\ln(N)$ values for 5000 Gaussian
  random noise time series. The probability of detecting a spurious
  peak with a power $P>6+\ln(N)$ is less than 0.36\%.}
\label{fig:noisehist}
\end{figure}

In order to do this, we simulated 5000 time series, randomly
sampling pure Gaussian noise with 0 mean and $\sigma=1$. 
Each time series consisted of $N$ data points,
with $N$ randomly chosen between 10 and 1000 using a uniform
distribution in log space.  We then applied the standard LS
analysis to each time series by searching through frequencies from
$2\pi$ to $N \pi$, and examined the distribution of $P_{\rm max}-\ln
(N)$ values, where $P_{\rm max}$ is the power corresponding to the
highest peak in each time series.  We found that $P_{\rm max}-\ln(N)$
tended to increase as we increased $N_{\omega}$ but above
$N_{\omega}=16 N$ the distributions remained virtually unchanged.
This worst case scenario is shown in Fig. \ref{fig:noisehist}.  Based
on this simulation, if we choose $P_{\rm cut}=6 +\ln(N)$ as our
significance criterion, then only 18 out of 5000 time series register
$P_{\rm max}$ above this threshold.  This corresponds to the false alarm
probability of $18/5000=0.0036$, which compares well to $p_f=0.0025$, which
we get from Eq. \eqref{eq:Pmin} if we set $N_{\omega} = N$.  We
adopted $P_{\rm cut} = \ln(N) + 6$ as our detection threshold,
i.e. only the peaks with power exceeding $P_{\rm cut}$ were treated as
real detections.

\section{The standard system}
\label{sec:stdsys}
Having established our method for timing analysis, we first test it
for what we call our standard binary lensing system.  It consists
of two $0.5 M_\sun$ stars separated by a distance of $2a = 2\,{\rm AU}$.
We would expect this type of binary system to be fairly common since
$0.5 M_\sun$ stars lie near the peak
of the IMF.  We take the source to be near the
Galactic Center at a distance of 8\,kpc and set the lens at
4\,kpc. For the relative transverse velocity, we adopt the value
$v=60$\,km/s.  For these parameters, the ratio of the binary
separation to the Einstein radius is $\alpha=0.25$ and the timescale
ratio is ${\cal R} = 0.31$.  These are fairly conservative choices.
On the one hand, $\alpha$ is low enough that the signatures of
binarity will not be very prominent, as demonstrated in
Figs. \ref{fig:samplelc} and \ref{fig:proretromaglc} \citep[see also
the discussion in][]{DSE14}.  In addition, a relatively low value of the
timescale ratio puts this binary firmly in the regime where both
the orbital motion and the transverse motion of the source will play 
important roles in the formation of the light curve, making it an ideal 
testing system for our timing analysis.  

\begin{figure}[h]
\begin{center}
\includegraphics[width=2.5in]{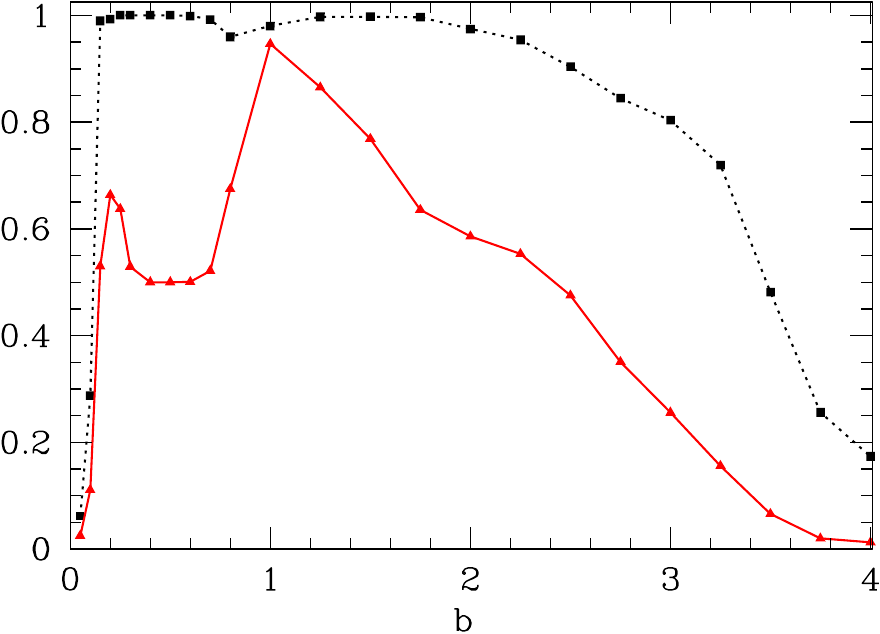}
\caption{The efficiency of our timing analysis for the
  standard system.  The dashed black line shows the period detection
  rate, i.e. the fraction of the simulated light curves for which of a
  significant periodic signal is detected.  The solid red line shows
  the correct period detection rate, i.e. the probability of
  extracting a significant orbital period that is within 10\% of the
  true period of 372.5\,days.}
\label{fig:stdcomb}
\end{center}
\end{figure}

Our goal is to investigate the success of our proposed period
extraction method for this system as a function of the impact
parameter $b$.  We want to determine how weak must be the overall
amplification of a microlensing event must be for the periodicity in the
light curve to no longer be detected.  To this effect, we vary the
value of $b$ from $0.05$ to $4.0$.  Since low $b$ means stronger events
and potentially more identified light curves in the existing lensing archives, 
we adopt a finer mesh for $b < 0.8$ and a coarser mesh for larger values of
$b$. 

For each value of $b$ we generated 2000 light curves equally split
between prograde and retrograde rotation directions. For each
light curve, the initial phase angle $\varphi_0$ was randomly chosen from the
interval $[0, 2\pi]$.  Each light curve was then analyzed using our
method to determine (1) whether the power spectrum contains at least
one peak with $P > P_{\rm cut}$, as defined in the previous section;
and (2) whether the period corresponding to the peak with the highest
power falls within 10\% of the true orbital period of the lensing
binary.

\subsection{Period Detection Rates}
\label{sec:detection}

The results are summarized in Fig. \ref{fig:stdcomb}.  As expected,
the overall period detection rate (shown as a dotted black line) falls
off with increasing $b$.  This makes perfect sense because as the
lensing event becomes weaker, the binarity signal becomes buried in
the noise.  Nevertheless, it is significant that most light curves are
identified as periodic past $b=3$.

\begin{figure}[h]
\begin{center}
\includegraphics[width=2.5in]{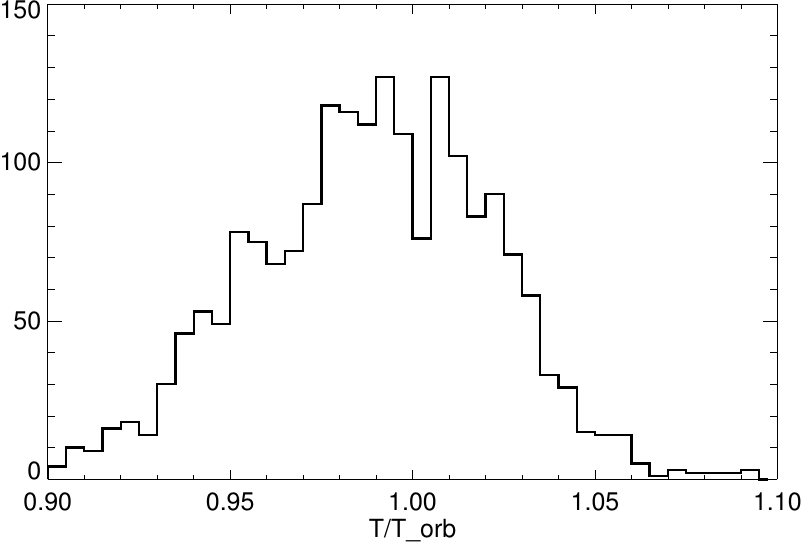}
\caption{Distribution of periods detected for $b=1.0$.}
\label{fig:periodhist}
\end{center}
\end{figure}

Of greater interest to us is the {\it correct} period detection rate
(shown as a solid red line), i.e. the probability of extracting a
significant orbital period that is reasonably close to the true period
of the binary.  For the purposes of this paper, we defined a detected
period to be ``reasonably close'' if it lies within 10\% of the true
binary period of the lensing system (this also includes having the
correct sign, i.e. distinguishing between prograde and retrograde
rotation).  This rate is falling faster with increasing $b$ than the
overall detection rate, but even at $b=3$, we are still correctly
identifying periods for 30\% of all the light curves.  Figure
\ref{fig:periodhist} shows the distribution of detected periods for
light curves with $b=1.0$.  It is clear that for most of these events we can
use our method to determine the binary period to an accuracy of a few
percent.

While the behavior of the detection rates for $b>1.0$ is easily
understood, there are two peculiar features evident for the
light curves with $b< 1.0$.  Firstly, for $b <0.2$ there is a steep
drop in both detection rates.  In addition, in the region $0,2 < b<
1.0$ there is a $\sim 50\%$ drop in the correct detection rate, while the 
overall detection rate remains near 100\%.  We investigate these in the
remainder of this section.

\subsection{High-frequency Signal in Low-$b$ Events}

\begin{figure}[h]
\begin{center}
\includegraphics[width=3.1in]{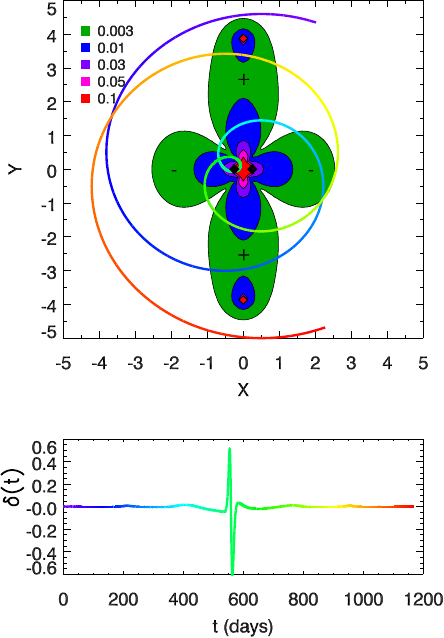}
\caption{Source trajectory and the corresponding
  residual light curve for a prograde standard system with $b=0.1$.}
\label{fig:b01map}
\end{center}
\end{figure}

To diagnose the precipitous drop in the detection rates for $b<0.2$,
we examine the trajectory and the 
corresponding residual light curve curves of an event with $b=0.1$ (see
Fig. \ref{fig:b01map}).

\begin{figure}[h]
\begin{center}
\includegraphics[width=3.0in]{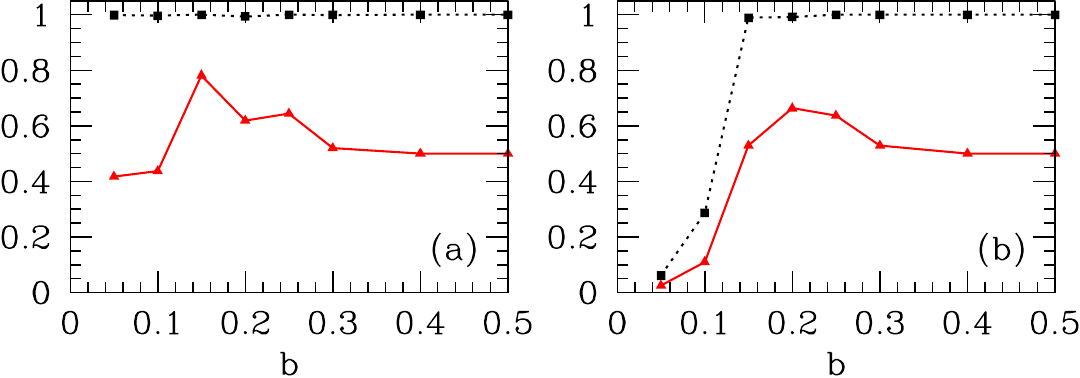}
\caption{Comparison of the detection rates with (panel [a]) and
  without (panel [b]) removing the high magnification ($A\ge 5$)
  portion of the light curves.  The dotted black and solid red curves
  are defined as in Fig. \ref{fig:stdcomb}.} 
\label{fig:newstd}
\end{center}
\end{figure}
The problem arises because the overall shape of the residual curve (lower panel of Fig. \ref{fig:b01map})
shows a single strong high-frequency oscillation near the peak of the
microlensing event. By relating the residual curve to the trajectory (upper panel of Fig. \ref{fig:b01map}), we
can see that it is due to the crossing of the red region between the
two binary companions.  Because of its relatively large amplitude,
this high-frequency feature dominates our timing analysis and no significant period is detected in the data
since it is never repeated .

A simple way of dealing with this issue is to simply remove the
high-frequency part of the light curve so that the global low-frequency
signal is given more weight in the timing analysis.  To determine
which part of the light curve to remove, we estimated the average
radius of the red region to be $\sim 0.2 \re$, and removed the
corresponding portion of the light curve, i.e. the region with the
overall magnification higher than $A_{\rm SL}(0.2)\approx 5$, where
$A_{\rm SL
}$ is given by Eq. (\ref{eq:amp}).  The detection rates with
and without this modification to the timing analysis are shown in
Fig. \ref{fig:newstd}.  Cutting out high magnification points appears
to be highly effective; it restores the period detection rate to near
100\% and raises the correct detection rate to at worst $50\%$
(Fig. \ref{fig:newstd}(a)).  Our investigation of other systems
(Section \ref{sec:othersys}) showed that his high-frequency signal
contamination during close-approach events appears to be an universal
problem.  Thus, any results we show from this point on include this
step in the timing analysis.

\subsection{Prograde--Retrograde Period Confusion}
\label{sec:prograde}
\begin{figure}
\begin{center}
\includegraphics[width=1.6in]{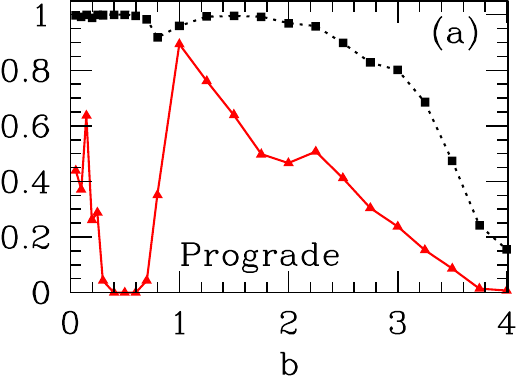} \hskip 0.3cm 
\includegraphics[width=1.6in]{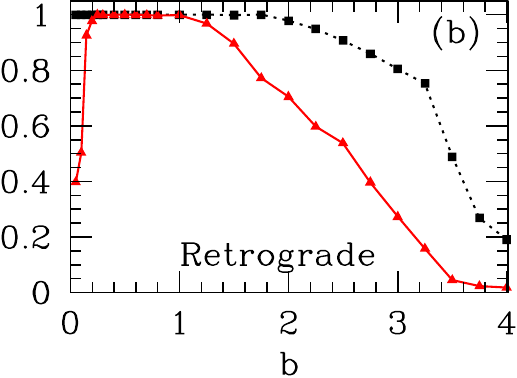}
\caption{Detection efficiency for the standard system
for light curves with prograde (panel (a)) and retrograde (panel (b))
rotation.}
\label{fig:stdproretro}
\end{center}
\end{figure}

To investigate the drop in the correct detection rate in the 
region  $0.2<b<1.0$, we examined the
statistics for prograde and retrograde systems separately.
Fig. \ref{fig:stdproretro} demonstrates that this effect 
is mainly due to the prograde systems.

\begin{figure}
\begin{center}
\includegraphics[width=3in]{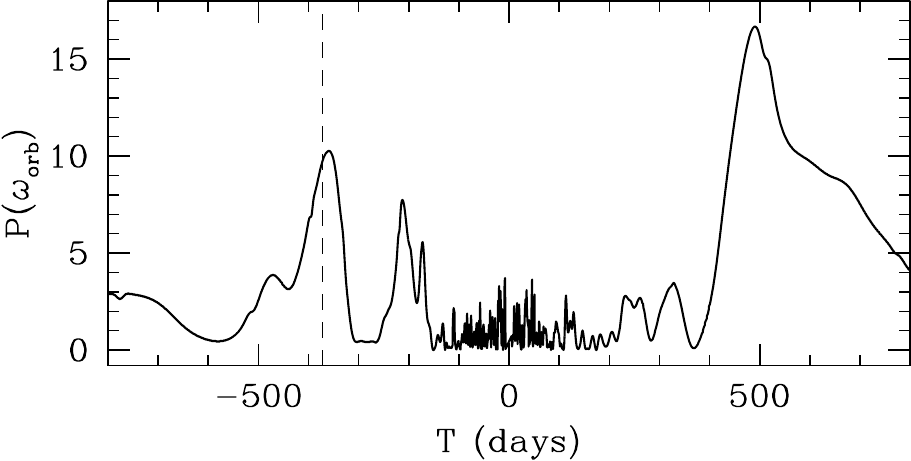}
\caption{Periodogram of a prograde standard system with $b=0.4$. The
  highest peak corresponds to a longer retrograde period of $+491$\,days
  while the true prograde period of $-372.5$\,days has a lower power. The
  dashed line shows the location of the true system period.}
\label{fig:b04powerspectrum}
\end{center}
\end{figure}

Next we examined the statistically significant though incorrect
periods detected for the prograde light curves.  It turns out that for
many of these events the highest peak in the periodograms 
corresponds to a longer retrograde period $\sim +500$\,days instead of
the correct prograde period of $-372.5$\,days, The correct period 
is also detected, but at a slightly lower
power than the retrograde period.  A periodogram of such 
a prograde light curve is shown in Fig. \ref{fig:b04powerspectrum}.

\begin{figure}
\begin{center}
\includegraphics[width=3in]{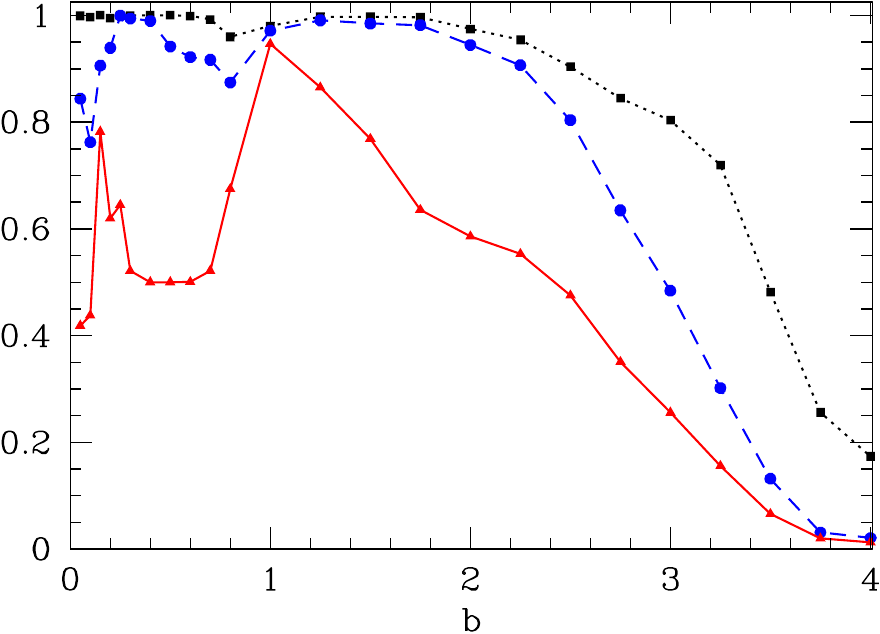}
\caption{Detection efficiency in standard system. The blue line
  represents the probability of either one of the highest peak and
  second highest peak lie in within 10\% of the correct period of the
  system. }
\label{fig:blueline}
\end{center}
\end{figure}

This confusion originates when subtracting the fitted single-lens
light curve.  Even a slight discrepancy between the actual and fitted
values for $b$, $\te$ and $t_0$ introduces an extra bump near the peak
of the light curve.  This extra feature causes the prograde residual
light curves to be mistaken for a retrograde one because the
introduction of the extra peak in the prograde case will mimic the
effect of higher frequency oscillations characteristic of retrograde
light curves.  In Fig. \ref{fig:blueline} we plot the probability that
one of the {\it two} highest-power peaks lies within 10\% of the
correct period (dashed blue line), in addition to the two detection
rates we have been discussing so far.  This new detection rate shows that 
indeed most of the missing periods for low-$b$ light curves appear in the 
timing analysis as second-highest-power peaks.  Interestingly, the detection
rate of periods for higher $b$ light curves also significantly improves
if we are willing to settle for two possible answers for the binary period. 

\begin{figure}[h]
\begin{center}
\includegraphics[width=3.3in]{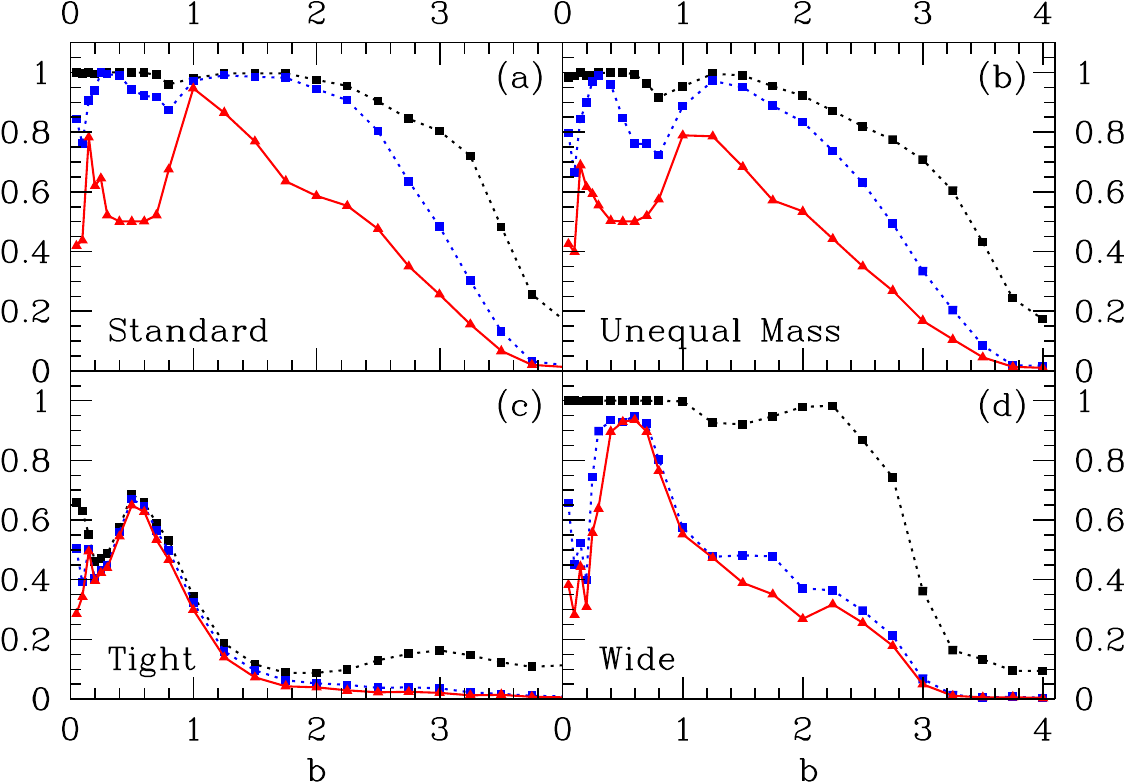}
\end{center}
\caption{Period detection rates for four systems summarized in Table 1. }
\label{fig:6sys}
\end{figure}

It turns out that this confusion between prograde and
retrograde rotation is specific to our
standard system rather than universal to binary lensing systems. In
most of the other example systems that we discuss in Section \ref{sec:othersys},
this issue hardly arises, except for one that is
most similar to the standard binary{, i.e. the probability 
of the highest peak being the correct period (red curve in Figure \ref{fig:6sys}) is the same 
as one of the highest two peaks being the correct period (blue curve in Figure \ref{fig:6sys})
for systems other than the standard and unequal.}  

\section{Other example systems}
\label{sec:othersys}
We now examine the efficiency of our timing analysis technique for
four other example lensing systems.  For all of them we take $D_{\rm l}$ and
$D_{\rm s}$ to be the same as for our standard binary.  The rest of their
physical parameters are summarized in Table \ref{table:6sys} and the
resulting period detection rates are plotted in Fig. \ref{fig:6sys}.
Note that panel (a) shows the results for our standard system for ease
of comparison.

\subsection{Unequal Mass Binary}
\begin{figure*}[tbp]
\begin{center}
\includegraphics[width=5.3in]{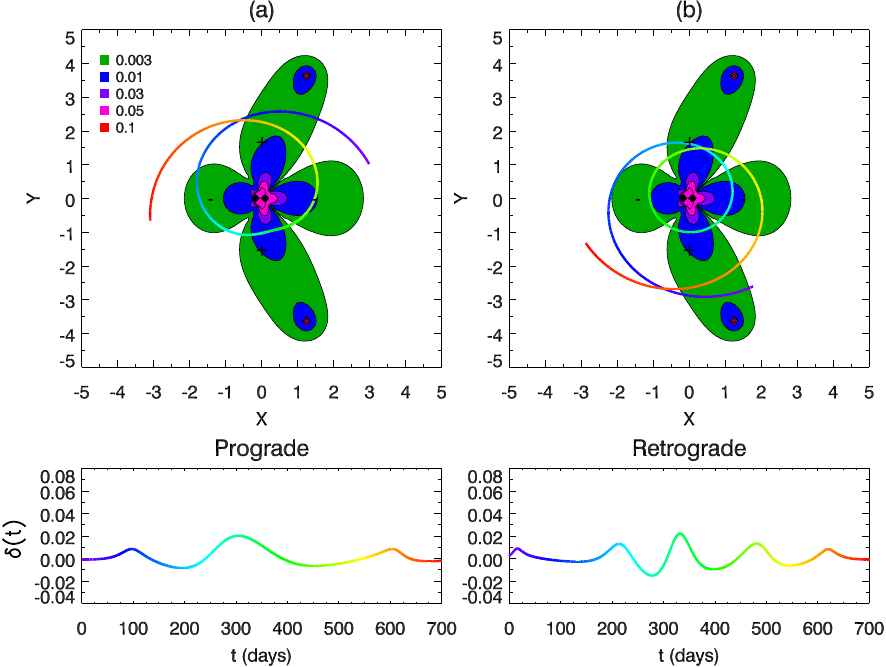}
\end{center}
\caption{Magnification map for the unequal-mass binary system.
  Since the two masses are not the same, the magnification map is not
  symmetrical with respect to the y-axis.  Also plotted are the source
  trajectories with $b=1.0$ for both prograde (panel (a)) and retrograde
  (panel (b)) binary rotation directions, as well as the corresponding
  residual light curves.}
\label{fig:unequalmaglc}
\end{figure*}

This binary system is essentially equivalent to the standard binary
lens system we discussed in the previous section except the masses of
the two companions are $\frac{1}{3}M_{\sun}$ and
$\frac{2}{3}M_{\sun}$, respectively.  The effect of having $q \neq 1$
is the change in the shape of the residual amplification map; as
illustrated in Fig. \ref{fig:unequalmaglc}, the spatial asymmetry of
mass distribution about the $y$-axis leads to the tilting of the
``petals'' in the residual pattern.  As the source follows its
spiraling trajectory in the lens plane, this asymmetry will certainly
affect the spacing of the peaks, as illustrated in the light curves 
shown in the lower panels in Fig. 
\ref{fig:unequalmaglc}, and can complicate the process of
extracting the orbital period.  However, it is encouraging that the
detection efficiency for this system, shown in Fig. \ref{fig:6sys}(b),
is only marginally different than for the standard equal-mass binary
(Fig. \ref{fig:6sys}[a]).  In fact, the overall shapes of the detection
rates for the two cases are essentially identical when $b < 1$. At
larger values of $b$, for which the asymmetry is more pronounced, the
unequal-mass system shows slightly lower detection rates, but the
drop does not exceed 5-10\%.

\subsection{Tight Binary}
\label{sec:tight}

This system differs from the standard case only in its smaller binary
separation.  With $\alpha=0.15$, 
the characteristic binarity features in the residual light curves
have very low amplitudes \citep[see e.g.][]{DSE14}.  On the other hand, smaller binary 
separation also means smaller orbital period. As a result, the timescale 
ratio is more than twice as large for this system as for our standard case, 
so the periodicity is easier to detect even in shorter duration light curves.

\begin{table}[tbh]
\begin{center}
\begin{tabular}{c|c|c|c|c|c|c}
\hline
System & $M/M_{\odot}$& $q$ & $\Torb$ (day)& $v$ (km/s)& $\alpha$ & ${\cal R}$\\
\hline
\hline
Standard&	1 &1 & 372.54	& 60 & 0.25 & 0.3135	\\
Unequal	&1& 0.5 &	372.54	& 60&	0.25	&	0.3135	\\
Tight	&1& 1 &	172.76	&	60 &0.15	&	0.676	\\
Wide	& 1& 1&	634.73	&	60 &0.36	&	0.184		\\
\end{tabular}
\caption{Summary of relevant parameters for the four example systems.}
\label{table:6sys}
\end{center}
\end{table}

Looking at the detection rates shown in Fig. \ref{fig:6sys}(c), it is
clear that our results are dominated by the decrease in $\alpha$, the
parameter which determines the amplitude of the features in the
residual light curves.  At best, we can detect periodicity in $\sim
60\%$ of all light curves and the probability of detection drops nearly
to zero for $b>1.5$.  However, because of the relatively high value of
${\cal R}$, when a period is detected the chances of it being correct
are very high.  The only exceptions are light curves with $b<0.1$,
where the correct detection rate drops to about half of the total
rate, just like it does for the standard system.  We show in
Section \ref{sec:lownoise} that the detection rates for this system are very
sensitive to the photometric precision of the data; with smaller
noise the detectability can become very high (see
Fig. \ref{fig:lownoise}(c)).

\subsection{Wide Binary}
\label{sec:wide}

We now consider the effect of having a larger binary separation while
keeping all the other parameters the same.  With $\alpha=0.36$, this
lensing system will produce light curves characterized by much more
pronounced deviations from the single-lens form. 
However, larger $a$ implies longer $T_{\rm orb}$ and therefore smaller 
${\cal R}$, making period detection more difficult.  

The results for the detection rates are shown in
Fig. \ref{fig:6sys}(d).  In contrast to the small-separation system
discussed in the section above, the overall detection rate for the
wide binary light curves shows only a moderate decrease compared to the
standard case.  However, the correct detection rate drops significantly in 
the regime where $b\gtrsim 1$, 
reaching a maximum of $50\%$ near $b=1.0$ and dropping off to zero 
for $b>3$.   

If we investigate the prograde and retrograde light curves separately,
we can see that the gap between the two detection rates in the $b\gtrsim1$ regime is caused
predominantly by the prograde systems.  In fact, taken by themselves,
prograde light curves have the correct detection rate close to zero.
The reason for this is simple.  At ${\cal R} = 0.184$ this binary 
straddles the boundary of detectability since its orbital period is
slightly longer than the length of the microlensing event.  In prograde 
binaries the effective period of the residuals is larger than $\Torb$, so 
detection becomes virtually impossible.  For retrograde binaries, the 
effective period is smaller than $\Torb$ and can therefore be still 
detected. The correct detection rate in the regime $b<1$
is less affected since the magnification map has more features 
near the center of the mass of the binary lens and thus 
effectively produce more periodic signals in the light curves, 
which eases the correct detection of the period.

\section{Low Noise Detection Rates}
\label{sec:lownoise}
\begin{figure}[h]
\begin{center}
\includegraphics[width=3.3in]{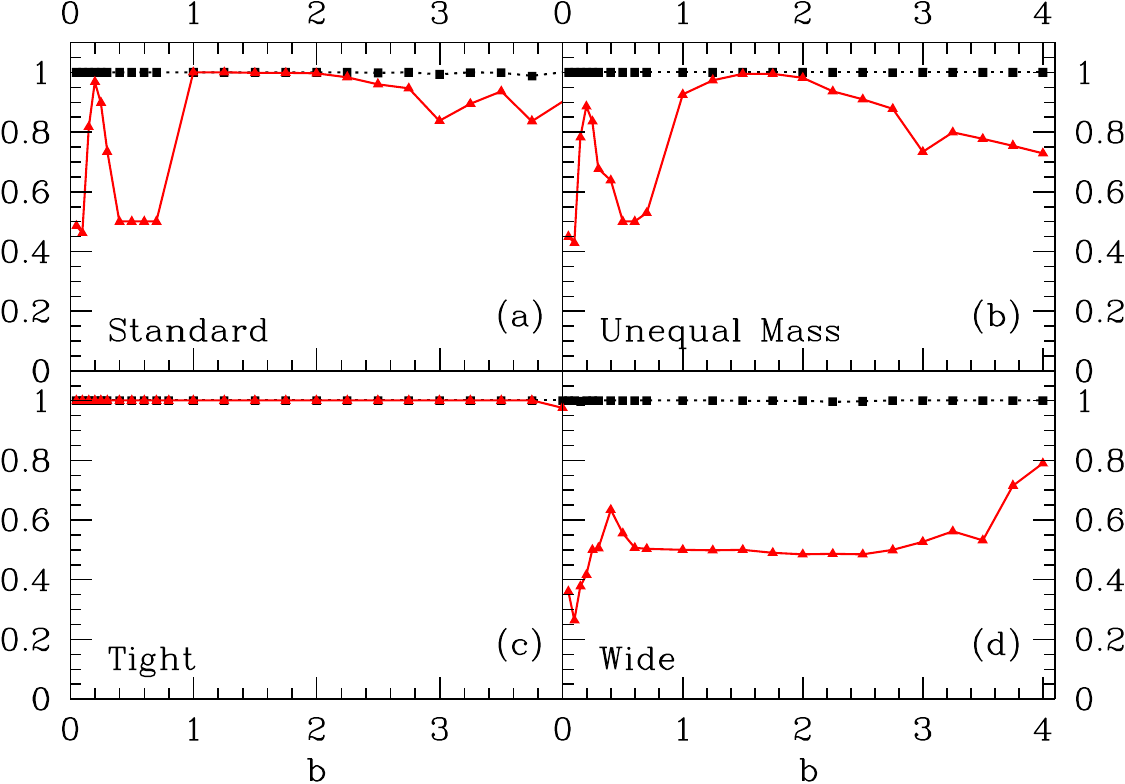}
\end{center}
\caption{Period detection rates for our four systems with 0.001\% photometric
uncertainties. The dashed black line shows the period detection rate
and the solid red line shows the correct
period detection rate as defined in Fig. \ref{fig:stdcomb}.
}
\label{fig:lownoise}
\end{figure}

While monitoring observations from the ground are unlikely to have
photometric errors smaller than $\sim 1\%$, space-based missions, such
as {\it Kepler}, {\it TESS}, and {\it WFIRST}, can do roughly 1000 times better \citep{Kepler,TESS}. It is highly
likely that multiple microlensing events will occur in the
field of those space-based telescopes over the lifetime of the mission, so it is worthwhile
to consider how our period detection method would fare for data with
photometric precision on the order of $0.001\%$.  To that end, we have
computed the period detection rates for our four systems in this low
noise limit.  Since high precision photometry would allow us to detect
much smaller deviations, we expanded the length of our simulated
light curves to $14 \te$, instead of $6 \te$ we adopted for higher
noise levels.

Fig. \ref{fig:lownoise} shows our new detection rates for the four
systems.  It is clear that the main effect of increasing photometric
precision is a significant increase in detection rates, especially at
high $b$.  This result makes perfect sense: at large $b$, the overall
light curve amplification as well as deviations from the point-lens
form become small enough as to be non-detectable from the ground, but
perfectly discernible from space.  

Note that the overall rate of period detection now remains flat at
essentially 100\% all the way beyond $b=4$.  Examining retrograde and
prograde light curves separately, we again find that most of the gap
between the overall and correct detection rates is attributable to
prograde rotation.  

The systems affected the most by the change in noise levels are the
tight and wide binaries.  The results for the former are particularly
striking; where before we could detect at most $60\%$ of the periods,
we now can detect the correct period for any light curve.  The main
reason for this improvement is again entirely due to the fact that 
very small amplitude signal can now be clearly detected.  The value of 
$\alpha$ for this system places the binary features in the microlensing 
light curves at the edge of detectability with 1\% errorbars, but this 
is not the case with improved precision.
\begin{figure}[h]
\begin{center}
\includegraphics[width=3.3in]{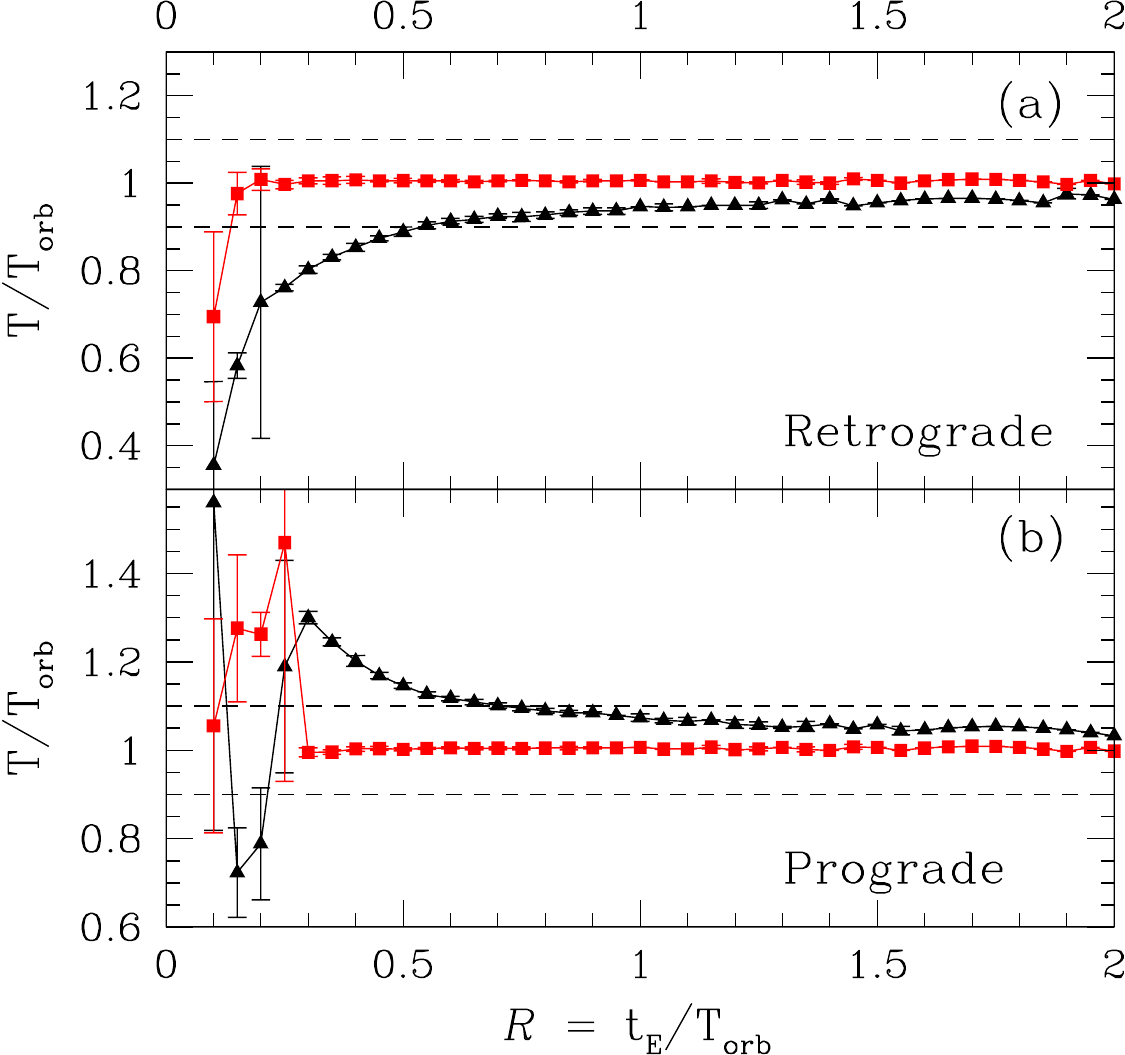}
\end{center}
\caption{Ratio of the detected to actual orbital periods of the
  binary lens as a function of ${\cal R}$ for LS (black triangles) and
  modified LS (red squares) periodogram methods.  The results for the
  former were divided by a factor of 2 to compensate for the symmetry
  of the amplification pattern.  Top and bottom panels show the
  results for retrograde and prograde light curves, respectively.}
\label{fig:modlscompare}
\end{figure}

For the wide binary, the periods for prograde light curves were largely
not detectable with higher noise levels since the length of the event 
was simply too long compared with the orbital period of the lens.  Improved 
sensitivity allows us to follow the event for a much longer time, effectively
decreasing the minimum value of ${\cal R}$ for which the periods can be 
reliably detected.\footnote{We point out that the drop of the correct detection rate
in the $b<1$ regime compared to the high-noise case (Fig. \ref{fig:6sys}(d))
is due to the prograde-retrograde period confusion explained in Section \ref{sec:prograde}. 
The correct detection rate taking into account the second peak in the 
periodogram is near $90\%$. This problem did not arise in the high-noise scenario
because the confused period is longer than $6t_E$ and was thus not 
searched for in the analysis of high-noise light curves.}

While the minimum value of ${\cal R}$ for which our timing analysis can be 
applied is set by the photometric precision of the data, we can place 
a firmer limit on the upper value of ${\cal R}$ for which our method is 
important.  Clearly, when ${\cal R} \gg 1$,
the lens-source relative motion becomes irrelevant and standard LS analysis 
should yield the correct period (up to a factor of 2) for such a binary lens.
To see when the standard LS results begin to deviate significantly from the actual orbital period, 
we analyze, using both our modified LS
and the standard LS methods, a series of 
light curves all with $b=1$, produced by lensing binaries with ${\cal R}$ 
spanning the range between $0.1$ and $2$. The other parameters of 
the lens binaries are the same as for our standard system with the exception of $a$, which 
was adjusted to vary $\Torb$ and therefore ${\cal R}$. 
Fig. \ref{fig:modlscompare} compares the ratio of detected orbital period to the true 
orbital period of the binary $T/T_{\rm orb}$ using our modified LS method (red curve)
and the standard LS method (black curve) as a function of ${\cal R}$.

We see that our proposed timing analysis
method is very successful in extracting the correct orbital periods for
the entire range of timescale ratios above the minimum set by the length 
of the light curves\footnote{Note that for high
  values of ${\cal R}$, $\alpha$ becomes very small, but the periods
  are still detectable with the low-noise data.  This would not be the
  case for 1\% photometry; however, other lensing parameters can be
  adjusted to ensure that the value of $\alpha$ remains within a
  detectable range \citep[see][]{DSE14}.  So our conclusions for ${\cal R}$
are still applicable.}, while the standard LS periodogram algorithm clearly begins
to fail (i.e. the detected period differs from the actual value by
more than $10\%$) when ${\cal R} \lesssim 0.8$. 

\section{Discussion and Conclusions}

When the Einstein crossing time is comparable to the orbital period of
a binary lens, microlensing light curves display quasi-periodic
features which can be used to determine $\Torb$.  In the limit when
${\cal R} \gtrsim 1$, the standard LS periodogram analysis can be used to
find the period.  However, in this paper we focus on the regime with
${\cal R} \lesssim 1$ when the variability timescale in observed
light curves is set by a combination of binary rotation and relative
source-lens motion.  We have shown that this regime includes specific
systems with orbital periods in the range from months to years.  For
such light curves the standard periodogram yields incorrect periods
(see Fig. \ref{fig:oldnewpowerspectrum}).  
{\cite{Nucita2014} have noticed
the same issue that the relative source-lens motion affects the periodic signals in 
binary microlensing light curves and thus 
prevents the correct orbital periodic extraction 
via the standard LS analysis. 
They proposed to circumvent the problem by removing the central part of the 
light curve and demonstrated its success on a few light curves.}
\footnote{It remains to be explored how sensitively the data-removal method depends on 
the configuration of the source-lens system, e.g. the impact parameter and the initial phase angle, and 
the noise level and sampling frequency under realistic observational conditions.}
In this paper, we proposed a modification to the standard LS timing analysis method designed to compensate for
the source motion and extract the correct orbital period.
We tested our new timing analysis method on simulated light curves {at 
two different noise levels} for four 
different binary lens systems and calculated period detection rates for a 
wide range of impact parameter values {averaged over random initial phase angles}.  The results are very encouraging.  
As long as the orbital period is detectable (i.e. not longer than the length
of the simulated light curve) our proposed method finds and identifies 
it correctly in a large fraction of cases.  

We find that the period detection rates are determined primarily by
the photometric precision and 
three interdependent parameters, $\alpha$, ${\cal R}$ and $b$.
Perhaps unsurprisingly, the photometric precision is the most 
important.  It determines the minimum value of $\alpha$ for which binarity 
of the lens is detectable \citep{DSE14} and sets the maximum value of $b$
for which a microlensing event is distinguishable from the noise.  
It also essentially sets the maximum practical length for a light curve 
which in turn determines the maximum detectable orbital period and therefore
places a lower limit on  ${\cal R}$.

We consider two levels of photometric uncertainty: 1\%, characteristic of 
ground-based data,  and 0.001\%, relevant for dedicated space missions such as 
\textit{Kepler}.  In the first case, the reasonable event duration is $6 \te$ (beyond
this, the wings disappear into the noise) and, 
correspondingly, the detection rate drops off for ${\cal R} \lesssim 1/6$,
as illustrated in Fig. \ref{fig:modlscompare} and well as by our results for 
the wide binary lens in Section \ref{sec:wide}.
High-precision photometry allows us to significantly expand 
the parameter space for which the orbital periods can be reliably detected with
our method.

We note that in our study, we model the light curves by treating the lensed star as a point source. 
This is well motivated by the fact that, for all the systems we considered in this paper, 
the angular size of a typical source star with radius $\sim 1 R_\odot$ is only $\sim 0.06\%$
of the Einstein radius of the binary lens, negligible compared to any of the feature size in the magnification map. 
Under certain circumstances, e.g. when the source is a giant star located very close to the lens, finite-source effect can become important, reducing the amplitude and broadening the features in the residual curve. However, \cite{Penny2011b} showed that 
even in such extreme cases, the finite-source effect only affects a small portion of the binary features in the light curve. More importantly, since the finite-source effect does not significantly alter the timing of the periodic features, the result of the 
timing analysis shall not be affected, as long as the softened features are still detectable above the noise level.

As a proof-of-concept for our modified LS analysis, the results we discuss in this paper are based on the analysis of
light curves produced by binary lenses in face-on circular orbits.  We
are now in the process of extending our calculations to include
elliptical and inclined orbits.  In these more complex situations, the
projected binary separation, and with it the shape of the lensing
magnification pattern, will vary with time during the microlensing
event.  This effect will certainly affect the spacing of the features
in the residual light curves, possibly affecting the extraction of
the orbital period.  Our preliminary results indicate that the orbital period
should still be detectable in a large fraction of light curves, but the 
full report will be the subject of a follow-up paper {(M. Vick et al, in 
preparation)}.

Finally, we want to point out that in its present form, our timing
analysis method works best for binary lenses with $q$ not very
different from unity.  It is predicated on the assumption that the
oscillatory features repeat twice per orbit.  The fact that period
detectability does not change much from $q=1.0$ to $q=0.5$
(Fig. \ref{fig:6sys}[a] and [b]) shows that some asymmetry in the
magnification pattern can be easily tolerated.  We tested lower values
of $q$ for the unequal-mass system and found that the detectability
decreases by a factor $\sim 2$ for $q=0.25$ and drops below $20\%$ for
$q=0.1$.  It is clear that in the planetary regime (i.e., when $q \ll 1$)
a factor of 2 in Eqs. (\ref{eq:modCj}) and (\ref{eq:modSj}) does not
apply, since the magnification pattern is no longer even remotely
symmetrical \citep{DiStefano2012}.  Fortunately, such systems produce visibly
different light curve morphologies characterized by very spiky rather
than sinusoidal features \citep[see][for some examples of light curves
  produced by low-$q$ binary lenses]{DSE14}, and so can be flagged. We
are now working on ways to extend our method to such low-$q$ systems.


\acknowledgements We would like to thank Christopher Night for significant contributions to
an earlier version of this work {and the anonymous referee for constructive comments.}
This work was supported in part by support from NSF AST-1211843,
AST-0708924 and AST-0908878 and NASA NNX12AE39GAR-13243.01-A.

\bibliographystyle{apj}
\bibliography{mybib}
\end{document}